\documentclass[11pt,a4paper]{article}
\usepackage{amsmath,amssymb,mathrsfs,sectsty}
\allsectionsfont{\normalsize}
\newcommand{\nc}{\newcommand}
\nc{\balpha}{\bar{\alpha}}\nc{\btheta}{\bar{\theta}}
\nc{\cc}{\mathrm{c.c.}}\nc{\const}{\mathrm{const}}
\nc{\bz}{\bar{z}}\nc{\mcp}{\mathscr{P}}\nc{\bbeta}{\bar{\beta}}
\nc{\bmcy}{\bar{\mathcal{Y}}}\nc{\mcy}{\mathcal{Y}}
\nc{\bgamma}{\bar{\gamma}}\nc{\bnu}{\bar{\nu}}
\renewcommand{\Im}{\,\mathrm{Im}\,}\renewcommand{\Re}{\,\mathrm{Re}\,}
\nc{\btau}{\bar{\tau}}\nc{\bmu}{\bar{\mu}}
\nc{\bpi}{\bar{\pi}}\nc{\bkappa}{\bar{\kappa}}
\nc{\bsigma}{\bar{\sigma}}
\nc{\Tr}{\mathrm{tr}\,}
\nc{\hP}{\hat{P}}
\nc{\bxi}{\bar{\xi}}

\nc{\blambda}{\bar{\lambda}}\nc{\be}{\begin{equation}}
\nc{\ee}{\end{equation}}\nc{\bE}{\bar{E}}\nc{\ud}{\mathrm{d}}
\nc{\bpartial}{\bar{\partial}}
\nc{\mch}{\mathcal{H}}\nc{\mcs}{\mathcal{S}}
\nc{\mcm}{\mathcal{M}}\nc{\mcr}{\mathscr{R}}
\nc{\mmcp}{\mathscr{P}}
\nc{\bL}{\bar{L}}\nc{\bB}{\bar{B}}\nc{\tm}{\tilde{m}}
\nc{\btm}{\bar{\tilde{m}}}\nc{\bn}{\bar{n}}
\nc{\bchi}{\bar{\chi}}\nc{\frec}[2]{{\textstyle{\frac{#1}{#2}}}}
\nc{\bW}{\bar{W}}\nc{\hodge}{{*}}\nc{\pbs}{\mathbf{s}}
\nc{\pbh}{\mathbf{h}}\nc{\gdzie}[1]{{\lower.2em\hbox{\big|}}_{#1}}
\nc{\mywedge}{\,{\scriptstyle{\wedge}}\,}
\nc{\mfd}{\mathfrak{D}}
\nc{\mfm}{\mathfrak{M}}
\nc{\mfi}{\mathfrak{I}}
\nc{\mfj}{\mathfrak{J}}
\nc{\mfh}{\mathfrak{H}}
\nc{\mdd}{\mathfrak{d}}
\newcommand{\innt}{\int\limits}
\newcommand{\qed}{\hfill $\Box$}
\numberwithin{equation}{section}
\newtheorem{tw}{Proposition}

\newtheorem{uw}{Remark}[section]

\begin{document}

\noindent {\bf Asymptotic stability of vacuum twisting type II metrics}

\medskip

\begin{small}
\noindent W{\l}odzimierz Natorf\footnote{email: {\tt nat@fuw.edu.pl}}

\medskip

\noindent Department of Theoretical Physics,\\
Faculty of Mathematics and Physics,\\
Charles University, V Hole{\v s}ovi{\v c}k{\'a}ch 2,\\ 
180\,00 Prague 8, Czech Republic.\end{small}

\begin{abstract} \noindent We generalize the result of Luk{\'a}cs {\it
et al.} on asymptotic stability of the Schwarz\-schild metric with
respect to perturbations in the Robinson-Traut\-man class of metrics
to the case of Petrov type II twisting metrics, uder the condition of
asymptotic flatness at future null infinity.  The Bondi energy is
used as the Lyapunov functional and we prove that the ``final state''
of such metrics is the Kerr metric.  \end{abstract}
\date{}

\section*{Conventions and notation}
The notation used here is similar to that of  \cite{KR}, with few exceptions.
Partial derivatives are denoted by comma.  Whenever asymptotically
flat space-time is mentioned, it is understood as space-time that
admits a piece of future null infinity $R\times S_2$. The standard metric
of $S_2$ in stereographic coordinates is $2 P_S^{-2}\ud\xi\ud\bxi$,
where $P_S= 1 + \frac{1}{2}|\xi|^2$.  Any other function $P$ divided
by $P_S$ will be denoted by $\hP$.

\section{Introduction}
The purpose of this paper is to show the Lyapunov stability
of  metrics known as diverging, twisting Petrov type II
metrics. They can be expressed by a
null tetrad \cite{KR}
\be\label{alstetrad}
\begin{aligned}
g & = 2(\theta^0\theta^1 - \theta^2\theta^3)\ ,\\
\theta^0 &= \omega = \ud u + (L\ud\xi+ \cc)\ ,\\
\theta^1 &= H\omega + \ud r + (W\ud\xi + \cc)\ ,\quad\\
\theta^2 &= P^{-1}(r-i\Sigma)\ud\xi\ ,\\
 \theta^3 &= \overline{\theta^2}\ ,
\end{aligned}
\begin{aligned}
\omega&\mywedge\ud\omega =:\frac{2\Sigma}{iP^2}
\ud u\mywedge\ud\xi\mywedge\ud\bxi\ ,\\
H&=-r(\ln P)_{,u}-\frac{mr+M\Sigma}{r^2+\Sigma^2}+K/2\ ,\\
K &= 2P^2\mathrm{Re}[\partial(\bpartial\ln P-\bL_{,u})]\ ,\\
\partial &= \partial_{\xi}-L\partial_u\ ,\\
W&=-(r+i\Sigma)L_{,u}+i\partial\Sigma\ .\end{aligned}
\ee
The shear-free repeated
principal null direction $\partial_r$ of the Weyl
tensor is related to $\omega$ via $\omega = g(\partial_r)$.
$P$, $m$ and $L$ are, respectively, two real and one complex
functions independent of $r$. Einstein equations in vacuum reduce to
\begin{subequations}
\be P^{-3} M = \Im \partial^2 \bpartial^2 V\ ,\label{eeqM}
\ee 
\be \partial(m+iM)=3(m+iM)L_{,u}\ ,\label{m3} \ee 
\be
(P^{-3}(m+iM)-\partial^2 \bpartial^2 V)_{,u}
=-P|I|^2\ ,\label{eqmloss} \ee 
\end{subequations} 
where\cite{RR} 
$P = V_{,u}$ and $I = (\bpartial + (\bpartial\ln P - \bL_{,u}))
(\bpartial\ln P - \bL_{,u})$. Define $G = -\partial\ln P 
+ L_{,u}$.

Equation \eqref{eqmloss} can be interpreted as the Bondi mass-loss
formula\cite{NT}: if $g$ is asymptotically flat, $|I|^2$ is
proportional to the Bondi news function, and the quantity
differentiated with respect to $u$ on the l.h.s. of \eqref{eqmloss} is
related to the modified Bondi mass aspect---the Bondi energy is given
by\cite{NT} \be\label{BE} E = \frac{1}{4\pi}\int_{S_2}\left(
\frac{m}{\hP^3} - P_S^3\Re \partial^2\bpartial^2 V \right)\ .  \ee
Condition \eqref{eeqM} is compatible with \eqref{eqmloss}
 because the r.h.s of \eqref{eqmloss} is real.

Apart from the freedom of fixing the origin of $r$, the allowed
coordinate transformations are of the form
\begin{subequations}
\be\label{cfu}
u\mapsto u'= F(u,\xi,\bxi)\ ,
\ee
\be\label{cfxi}
\xi \mapsto \xi'=h(\xi),\ h_{,\bxi}= 0\ ,
\ee
\end{subequations}
Under \eqref{cfu}, $P\mapsto P' = (F_{,u})^{-1} P$ and
$L\mapsto L' = -\partial F$. In particular, by means
of \eqref{cfu} one can achieve $P_{,u}=0$. Then the
remaining freedom is \eqref{cfu} with $F_{,uu}=0$.
The function $G$ transforms as $G\mapsto G' = 
h'^{-1} G - h''/(2 h'^2)$ under \eqref{cfxi} and
is preserved by \eqref{cfu}. 

\bigskip

Asymptotic stability of the Schwarzschild solution within the
Robinson-Traut\-man (RT) class of metrics\cite{RT} has been established in
 \cite{LPPS}.  The Lyapunov functional used by Luk{\'a}cs {\it et
  al.}  was $\int_{\mcs} K^2$, where $\mcs=\{u=\const,\,r=\const\}$
  and $K$ is its Gauss curvature\footnote{In the published version
  this sentence is typeset in a way that makes no sense},
and they had to apply the Laplace operator of $\mcs$
($2P^2\partial_{\xi} \partial_{\bxi}$) to the Robinson-Trautman
equation to make it {\it an evolution of $K$ alone}. This seems rather
complicated as compared to noting that $\int \hP^{-3}$ also decreases
with $u$. The authors of  \cite{LPPS} most probably were unaware
that the average
$ \langle m \hP^{-3}\rangle_{u=\const}$  
is the Bondi mass of an asymptotically flat RT
space-time and that the RT equation is responsible for the energy
loss,  but still it is evident that for
positive, smooth $\hP$ and $m>0$ that average is positive and
that its $u$-derivative is negative (being the average of the news
function). Establishing this is attributed to D.~Singleton
in  \cite{CL}.

In the work presented here, we have no natural ``finite $r$'' and
``constant retarded time'' family of two-dimensional closed
surfaces. Our idea is to assume asymptotic flatness of $g$ and use the
Lyapunov functional which is identical with the Bondi energy (which we
also assume to be positive), $\mathcal{L}[g] = E$, where $E$ is given
by \eqref{BE}. Then the mass-loss equation \eqref{eqmloss} shows that
$ \frac{\ud}{\ud u} \mathcal{L}[g] \leq 0 $ and $\mathcal{L}[g]=\const
\Leftrightarrow I=0$. Therefore it is important to know the properties
of twisting metrics with vanishing news, and whether $I=0$ is
compatible with asymptotic flatness. We will show that the only
possible ``fixed point'' solution that is asymptotically flat is the
Kerr metric.

\section{``Non-radiating'' type II  metrics}
The condition $I=0$ is written as
\be\label{G}
0= \bar{I} = (G-\partial)G\ .
\ee
In the gauge $P=1$, $G = L_{,u}$ and this is integrated with the help
of the Cauchy-Riemann function, similarly as \eqref{m3}\cite{ST}. 
Suppose we have such a function $\chi$ that $\partial\bchi
= 0$. Then $L = \frac{\bchi_{,\xi}}{\bchi_{,u}}$, and due to
$[\partial_u,\partial] = -L_{,u}\partial_u$ we have $G = \bchi_{,u}$
and $m + i M = (\bchi_{,u})^3$, but such solution is defined up to
multiplication by a Cauchy-Riemann function.  Let us take, similarly
as in  \cite{KR}, $m + i M = 2 G^3 A(u,\xi,\bxi)$.  Imposing \eqref{m3}
we obtain $\partial A = 0$. But the field equation \eqref{eqmloss}
yields $(m+iM)_{,u} = 0$, so $G^3 A$ is $u$-independent. Applying
$\partial$ to $G^3 A$ we get 
\be\label{ga1} \partial(G^3 A) = (G^3
A)_{,\xi} = 3 A G^2\partial G = 3 G^4 A\ .\ee 
Taking $u$-derivatives of both sides of \eqref{ga1} gives
\be\label{ga2} 0 = (G^3 A)_{,\xi u} = 3[G(G^3 A)]_{,u} = 3 G_{,u} G^3
A + 3G (G^3 A)_{,u} = 3 G^3 A G_{,u}\ .\ee 
The conclusion is that $G_{,u} = 0$ is not an extra assumption as
stated in  \cite{KR}. Also, $A_{,u} = A_{,\xi} = 0$ and we replace $A$
by $\bar{a}(\bxi)$, an antiholomorphic function.  The condition
\eqref{G} gives\cite{KR} $G = -(\xi +
\bar{g}(\bxi))^{-1}$, where $g$ is holomorphic (not to be
misidentified with the metric).  \begin{uw} The function $G^{-1}$ is
harmonic.\end{uw}

Now, following  \cite{KR}, we consider separately the following
two subclasses of metrics with vanishing ``news'', 
for which $L_{,u}$ is either transformable to zero or not. 

\subsection{$I=0$, $L_{,u}$ transformable to $0$}\label{sol1}
In this case $\partial_u$ is a Killing
vector of $g$. The field equations give
\be\label{ll}
\begin{aligned}
P & = A\xi\bxi + (B\xi + \cc) + C\ ,\\
m + i M & = \bar{z}(\bxi)\ ,\\
L & = P^{-2}\left(\bar{l}(\bxi)-\frac{1}{2}\innt\frac{z(\xi)}
{(A\xi + \bar{B})^2}\ud\xi\right)\ .
\end{aligned}\ee
For $AC-|B|^2>0$, linear transformation of $\xi$ gives $P=P_S$. Note
that $m+iM$ cannot be regular unless $z=\const$ and $l$ is linear, and
the metric cannot be asymptotically flat unless $M=0$. The reason is
that $\xi^{-1}$ terms in front of $m$ can be absorbed in $u$ by means
of \eqref{cfu} with $F_{,uu}=0$, under which $L\mapsto L - F_{,\xi}$.
Similar terms in front of $M$ would survive since $F$ is
real. Therefore $g$ is a vacuum Kerr-Schild metric. According to
 \cite{KW}, the only metric of this kind with singularities 
inside a spatially
bounded region is the Kerr metric. (Another way to see this is
to use the Weyl scalars \cite{WK} and deduce that $I=0$ implies that
$g$ is of Petrov type D, and the only vacuum solution of this type
with a $S_2$ set of generators of future scri is the Kerr metric.)

\subsection{$I=0$, $L_{,u}$ not transformable to $0$}
In this case for $P_{,u} = 0$, say $P=P_S$, the field equations give
\be\label{lll}
\begin{aligned}
m + i M & = 2 P_S^3 G^3 \bar{a}(\bxi)\ ,\\
L & = (G + (\ln P)_{,\xi})u + P^{-1}G \ell(\xi,\bxi)\ ,\\
\ell & = - \innt a(\xi) \bar{G}(\bxi,\xi) G(\xi,\bxi)^{-1}\ud \xi
+ \bar{\varphi}(\bxi)\ ,
\end{aligned}\ee
where $a$ and $\varphi$ are arbitrary holomorphic functions of
$\xi$. 
To begin discussing regularity of the metric components, consider the
blow-up points of $G$, i.e. the set $\mathsf{N} =
\{\xi\in\mathbb{C}\colon\xi + \bar{g}(\bxi)=0\}$. From the harmonicity
of $G^{-1}$ it follows that $\mathsf{N}$ cannot be a two-dimensional
subset of $R^2$, so (if nonempty), it must be either a curve or a set
of points.

If $\mathsf{N}\neq\emptyset$, we can solve its defining equation. Let
$\xi = |\xi| e^{i\beta}$ and $g(\xi) = |g| e^{i\gamma}$, where $\beta$
and $\gamma$ are real phases. On $\mathsf{N}$ we have $|g| = |\xi|$
and since $\xi = - \bar{g}$, their phases satisfy $\pi - (\beta +
\gamma) = 0 \mod 2\pi$. This gives $g = -|\xi|e^{-i\beta}$, but $g$ is
holomorphic, so the only possibility is $|\xi| = \const$ and $\bar{g}
=\const \bxi^{-1}$. Therefore $\mathsf{N}$ is a circle or a point
(respectively, when $\const > 0$ or $\const = 0$).  In both cases we
cannot cancel such set of zeros of $G^{-1}$ by specifying the function
$\bar{a}$, because it is antiholomorphic.

The most interesting case is that of $\mathsf{N}=\emptyset$.
A family of $g$'s giving rise to regular $m+iM$ is given
by \be\label{galpha}
\overline{g_{(p,\alpha)}} =
|p|e^{i\alpha}\bxi^{-1},\ R\ni\alpha\neq 0\mod\pi\ .
\ee
\begin{uw} $P_S G_{(p,\alpha)}\bxi^{-1}$ is a regular function, so the
regularity of $m+iM$ can be achieved by taking $\bar{a} = \const\,
\bxi^{-3}$.  \end{uw} 
\begin{uw} Adding a linear function of $\bxi$ to
$\overline{g_{(p,\alpha)}}$ does not spoil asymptotics of $G$, but
always introduces nontrivial zeros in its denominator.  \end{uw}
This, together with harmonicity of $G^{-1}$ suggests that
$G_{(p,\alpha)}$ defined by \eqref{galpha} is the only
possible function $G$ that can give finite $m+iM$.
Note that for $\alpha\notin \pi Z$, the Newman-Unti-Taub (NUT) charge
 $M$ cannot vanish: $L_{,u}$ is not transformable
to zero, so $m +  i M$ cannot be made constant by
means of \eqref{cfu}, therefore $M\neq 0$, and ``scri'' would
have topology  $S_1\times S_2$ rather than $R\times S_2$.

A more direct argument showing singularity in $L$ is based on finding
$\ell \sim [ |\xi|^{-2} + 2 i |p|^{-1} (\sin\alpha) \ln(|\xi|^2 +
|p|^{-1} e^{i\alpha})]$ and noting that 
$P_S^{-1}G_{(p,\alpha)}\ell$ must be singular because of the
term $|\xi|^{-2}$, since $P_S G_{(p,\alpha)}\xi^{-1}$ is well
defined.

\section*{Summary}
\begin{tw} 
Regularity conditions imposed on the two subclasses
\eqref{ll} and \eqref{lll} 
of ``news-free'' metrics \eqref{alstetrad} rule out the second
subclass and single out the Kerr solution out of
the first subclass.\qed\end{tw}
\begin{tw} 
Suppose $g$ given by \eqref{alstetrad} is asymptotically flat
and equation \eqref{eqmloss} admits a solution that
 can be continued to $u = \infty$.
Then the Kerr metric is  asymptotically stable 
and $g$ tends to Kerr as $u\to \infty$. \qed
\end{tw}
Problem of the existence and regularity  of solutions of 
\eqref{eqmloss} as well as their approximations 
shall be treated elsewhere. 

Note that if one wishes to
treat these metrics as perturbations of the Schwarz\-schild metric,
a stationary perturbation leading to the
Kerr metric must be included. In other words: metrics of Petrov
type II with twist cannot be viewed as nontrivially evolving 
corrections to Robinson-Trautman metrics.

\section*{Acknowledgments}
I would like to thank Professors J.~Jezierski and E.~T.~Newman for
help with understanding the problems with NUT parameter and asymptotic
flatness in GR.

I am grateful to people at the Department of Theoretical Physics at
Charles University for warm hospitality during my visits to Prague.

This research is supported by grant No. 202/09/0772 of the Grant
Agency of Czech Republic.



\begin{thebibliography}{100}
\bibitem{KR} Stephani H., Kramer D., MacCallum M. A. H.,
Hoenselaers C. and Herlt E.,
{\it Exact Solutions to Einstein's Field Equations, Second Edition},
2003 (Cambridge: Cambridge University Press)

\bibitem{RR} Robinson I., Robinson J. R.,
{\it Vacuum metrics without symmetry,}
{\it Int. J. Theor. Phys.} \underline{7} (1969), 231.

\bibitem{NT} Natorf W., Tafel J. {\it Asymptotic
flatness and algebraically special metrics},  
Class. Quantum Grav. \underline{21} (2004), 5397.

\bibitem{RT} Robinson I., Trautman A., 
{\it Some spherical gravitational waves in
general relativity}, Proc. Roy. Soc. Lond. A \underline{265}
(1962), 463.

\bibitem{LPPS} Luk{\'a}cs B., Perj{\'e}s Z.,
Porter J., Sebesty{\'e}n {\'A}., {\it 
Lyapunov Functional Approach to Radiative Metrics},
Gen. Rel. Grav. \underline{16} (1984), 691.

\bibitem{CL} Chow E. W. M., Lun A. W.-C.,
 {\it Apparent Horizons in Vacuum Robinson-Trautman Spacetimes},
J. Austral. Math. Soc. Ser. B \underline{41} (1999), 217.

\bibitem{ST} Stephani H., {\it Algebraically
special, diverging vacuum and pure radiation fields
revisited}, Gen. Rel. Grav. \underline{16} (1983), 173.

\bibitem{KW} Kerr R.P., Wilson W.B. {\it Singularities in the
Kerr-Schild metrics}, Gen. Rel. Grav. \underline{10}  (1979), 273.

\bibitem{WK} G. J. Weir, Kerr R. P.,
{\it Diverging type-D metrics}, 
Proc. R. Soc. London, Ser. A \underline{355} (1977), 31. 

\end{thebibliography}
\end{document}